\documentclass[submission,copyright,creativecommons]{eptcs}
\providecommand{\event}{LFMTP 2019} 

\usepackage{underscore}
\usepackage{proof}
\usepackage{amssymb}
\usepackage{stmaryrd}
\usepackage{cedilleverbatim}

\mathchardef\mhyph="2D 

\newcommand{\interp}[1]{[\negthinspace[ #1 ]\negthinspace]}
\newcommand{\abs}[4]{{#1}\, #2\! : \! #3.\, #4}
\newcommand{\absu}[3]{{#1}\, #2.\, #3}
\newcommand{\lam}[2]{\lambda\, #1.\, #2}
\newcommand{\Lam}[2]{\Lambda\, #1.\, #2}
\newcommand{\tpcheck}[0]{\Leftarrow}
\newcommand{\tpsynth}[0]{\Rightarrow}

\newcommand{\authorrunning}[0]{A. Stump}
\newcommand{\titlerunning}[0]{Towards HOAS in Cedille}

\title{A Weakly Initial Algebra for Higher-Order Abstract Syntax in Cedille}
\author{Aaron Stump
\institute{Computer Science\\ The University of Iowa\\ Iowa City, Iowa, USA}
\email{aaron-stump@uiowa.edu}}

\begin{document}
\maketitle

\begin{abstract}
  Cedille is a relatively recent tool based on a Curry-style pure type
  theory, without a primitive datatype system.  Using novel techniques
  based on dependent intersection types, inductive datatypes with
  their induction principles are derived.  One benefit of this
  approach is that it allows exploration of new or advanced forms of
  inductive datatypes.  This paper reports work in progress on one
  such form, namely higher-order abstract syntax (HOAS).  We consider
  the nature of HOAS in the setting of pure type theory, comparing
  with the traditional concept of environment models for lambda
  calculus.  We see an alternative, based on what we term Kripke
  function-spaces, for which we can derive a weakly initial algebra in
  Cedille.  Several examples are given using the encoding.

\end{abstract}

\section{Introduction}

Modern constructive type theory is based on a decades-long development
of formal systems, culminating in current tools like Coq and Agda, to
name two of the most widely used~\cite{coq,agda}.  To summarize the
relevant history: in the 1980s Coquand and Huet proposed the Calculus
of Constructions (CC) as a synthesis of impredicative type theory as
independently proposed by Girard and
Reynolds~\cite{girard72,reynolds74}, and dependent type theory as
found in de Bruijn's Automath and further developed by
Martin-L\"of~\cite{debruijn80,martinlof73}.  What was initially
believed by researchers working on CC was confirmed in the early 2000s
by Geuvers: induction is not derivable in CC (although note that
technically, Geuvers's theorem is about just the second-order fragment
of CC)~\cite{geuvers01}.  So in the late 1980s and early 1990s,
researchers explored various ways of adding primitive inductive
datatypes to CC~\cite{mohring93,pfenning89}.  At the same time, Luo
analyzed an extension of CC with an $\omega$-indexed predicative
hierarchy of universes~\cite{luo90}, still found in Coq today.  A
practically viable solution to the problem of inductive datatypes was
reached in Werner's development of the Calculus of Inductive
Constructions (CIC), which added a specific class of inductive
datatypes to CC (note that the predicative hierarchy is not included
in CIC as analyzed by Werner)~\cite{werner94}.  Subsequent work on the
theory and practice of Coq has built upon these results, resulting in
a tool that is both widely used and rightly generally considered a
great success.

Despite these excellent achievements, there are two notable issues
with CIC's solution to the problem of datatypes in type theory:
\begin{enumerate}
\item The class of datatypes is fixed as part of the definition of the theory.
\item The core theory upon which the complex edifice of the rest of the proof
  assistant is built must include support for that class of inductive datatypes,
  as they are primitive to the theory.
\end{enumerate}
(1) is an issue because it means that subsequent discoveries and
proposals for advanced forms of datatypes are excluded from CIC.  One
would have to rework the entire metatheory of CIC to add them.  Or one
could adopt the approach taken in Agda, which is to extend the
datatype system without requiring full metatheoretic justification.
While this facilitates exploration of advanced forms of datatypes, it
comes at the risk of introducing inconsistency into the theory
(through a novel form of datatype that would turn out to be logically
unsound).  (2) is an issue because it means that the trusted computing
base of a tool like Coq is rather large.  At present, for example, the
kernel of Coq -- the internal code which one must trust when the
type-checker accepts a theorem (this does not count parsers and
printers; cf.~\cite{wiedijk12}) -- is just over 30k lines of OCaml.
This includes powerful features like byte-code compilation for faster
conversion-checking, which could be excluded from the line count just
for core typing; but even the files for inductive types
(\verb|indtypes.ml| and \verb|inductive.ml|) total just under 2200
lines (see \url{https://github.com/coq}).  It would be very nice to
have a core checker under, say, 1000 lines of functional code.

Cedille is a recently released proof assistant based on a novel
minimalistic extension of CC, which allows derivation of inductive
datatypes with their induction principles.  So the core theory does
not include a primitive notion of inductive datatype, and indeed can
be checked in under 1000 lines of Haskell~\cite{stump18c}. Cedille is
briefly described in Section~\ref{cedille}.  The focus of the current
paper is on work in progress deriving an advanced form of datatype in
Cedille, namely higher-order abstract syntax (HOAS)~\cite{pfenning88}.
Section~\ref{sec:hoas} discusses what HOAS should be taken to mean in
the context of pure lambda calculus (where every term is encoded
functionally), considering (and rejecting) the traditional environment
models for algebraic semantics of lambda calculus.
Section~\ref{sec:encoding} presents an alternative implemented in
Cedille, for which we have a weakly initial algebra.  This approach
uses what we term \emph{Kripke function spaces} to allow construction
of an encoded nested $\lambda$-abstraction.  It turns out that for
what has been achieved so far, the full power of Cedille is not needed,
and the code can also be written in Haskell with a few language extensions (Section~\ref{sec:hs}).
Section~\ref{sec:param} discusses a possible way to extend this to
obtain induction, based on parametricity.

\section{Cedille and its Type Theory}
\label{cedille}
We briefly summarize the type theory of Cedille, called the Calculus of Lambda
Eliminations (CDLE).  The system has evolved
from an initial version~\cite{stump17a}, to its current
form~\cite{stump18b}.  Several other works demonstrate applications of
the theory to derivation of inductive
datatypes~\cite{firsov18b,firsov18,stump18}, and to zero-cost
coercions between related datatypes~\cite{diehl18}. The main
metatheoretic property proved in previous work is logical consistency:
there are types which are not inhabited.  All the code appearing in
this paper can be checked using Cedille 1.0.  (Cedille 1.1 adds
datatypes which elaborate down to the pure type theory of CDLE, but we
do not make use of this feature here.)

CDLE is an extrinsic (i.e. Curry-style) type theory, whose terms are
exactly those of the pure untyped lambda calculus (with no additional
constants or constructs).  The type-assignment system for CDLE is not
subject-directed, and thus cannot be used directly as a typing
algorithm.  Indeed, since CDLE includes Curry-style System F as a
subsystem, type assignment is undecidable~\cite{Wells99}.  To obtain a
usable type theory, Cedille combines bidirectional
checking~\cite{pierce+00} with a system of annotations for terms, to
obtain algorithmic typing.  But true to the extrinsic nature of the
theory, these annotations play no computational role, and are erased
both during compilation and before formal reasoning about terms within
the type theory, in particular by definitional equality.  We summarize the central
rules and clauses of the erasure function in
Figure~\ref{cdle} and following text.  As this is, by necessity of space, quite
brief, please see a report for full details, including semantics and
soundness results~\cite{stump18b}.

\begin{figure}
\centering
\minipage[b]{0.75\textwidth}
  \[
  \begin{array}{ll}
    \infer{\Gamma\vdash \absu{\Lambda}{x}{t} \tpcheck \abs{\forall}{x}{T'}{T}}{\Gamma,x:T'\vdash t \tpcheck T & x\not\in\textit{FV}(|t|)} &
    \hspace*{5pt}\infer{\Gamma\vdash t\ \mhyph t' \tpsynth [t'/x]T}{\Gamma\vdash t \tpsynth \abs{\forall}{x}{T'}{T} & \Gamma\vdash t' \tpcheck T'} \\ \\

    \infer{\Gamma\vdash \beta\{t'\} \tpcheck \{ t \simeq t \}}{\Gamma\vdash \textit{FV}(t)\subseteq \textit{dom}(\Gamma)} &
    \infer{\Gamma\vdash \rho\ t'\ \mhyph\ t \Rightarrow [t_2/x]T}
          {\Gamma\vdash t' \tpsynth t_1 \simeq t_2 & \Gamma \vdash t \Leftrightarrow [t_1/x]T}
\\ \\

    \infer{\Gamma\vdash [ t , t' ] \tpcheck \abs{\iota}{x}{T}{T'}}
          {\Gamma\vdash t \tpcheck T & \Gamma\vdash t' \tpcheck [t/x]T' & |t| =_{\beta\eta} |t'|} &
    \infer{\Gamma\vdash t.1 \tpsynth T}{\Gamma\vdash t \tpsynth \abs{\iota}{x}{T}{T'}}
\\ \\
    \infer{\Gamma\vdash t.2 \tpsynth [t.1/x]T'}{\Gamma\vdash t \tpsynth \abs{\iota}{x}{T}{T'}} &
    \infer{\Gamma\vdash \chi\ T\ \mhyph\ t \tpcheck T'}
          {\Gamma\vdash T\tpcheck \star & \Gamma\vdash t \tpcheck T & T \cong T'}
\\ \\

    \infer{\Gamma\vdash \chi\ T\ \mhyph\ t \tpsynth T}
          {\Gamma\vdash T\tpcheck \star & \Gamma\vdash t \tpsynth T' & T \cong T'} &
    \infer{\Gamma\vdash \phi\ t\ \mhyph\ t'\{t''\} \Leftrightarrow T}
          {\Gamma\vdash t\tpsynth \{t'\simeq t''\} & \Gamma\vdash t' \Leftrightarrow T}

  \end{array}
  \]
\endminipage\hfill
\minipage[b]{0.25\textwidth}
\[
  \hspace*{5pt}\begin{array}{lll}
    |\abs{\Lambda}{x}{T}{t}| & = & |t| \\
    |t\ \mhyph t'| & = & |t| \\
    |\beta\{t\}| & = & |t|\\
    |\rho\ t'\ \mhyph \ t| & = & |t| \\
    |\phi\ q\ \mhyph \ t_1 \{t_2\}| & = & |t_2| \\
    |[t_1,t_2]| & = & |t_1| \\
    |t.1| & = & |t| \\
    |t.2| & = & |t|
  \end{array}
  \]

\endminipage
\caption{Introduction, elimination, and erasure rules for additional type constructs.  Note that $\Leftarrow$ is for checking mode, $\Rightarrow$ is for synthesizing, and $\Leftrightarrow$ refers to either mode.}
\label{cdle}
\end{figure}

CDLE extends the (Curry-style) Calculus of Constructions (CC) with
a primitive intensional untyped equality, intersection types, and implicit
products (in the following explanation we use \verb|tt| fonts to introduce
the concrete syntax, very close to the mathematical one, expected by Cedille):
\begin{itemize}
\item \verb;{ t₁ ≃ t₂ };, an intensional equality type between terms
  \verb;t₁; and \verb;t₂; which need not be typable at all.  We
  introduce this with a constant \verb;β{t}; which erases to
  erasure of \verb;t; (so our type-assignment system has no additional
  constants, as promised); \verb;β{t}; proves \verb;{ t' ≃ t' }; for any
  term \verb;t'; with free variables all in scope.  Combined with definitional equality,
  \verb;β{t}; proves \verb;{ t₁ ≃ t₂ }; for any $\beta\eta$-equal \verb;t₁;
  and \verb;t₂; whose free variables are all declared in the typing
  context.  If the term \verb|t| is omitted from \verb;β{t};, then it is assumed
  to be \verb;λ x. x;. We eliminate the equality type by rewriting, with a
  construct \verb;ρ t' - t;.  Suppose \verb;t'; proves \verb;{ t₁ ≃ t₂ };
  and we are checking the $\rho$-term against a type \verb;T;, where
  \verb;T; has several occurrences of terms definitionally equal to
  \verb;t₁;.  Then bidirectional typing proceeds by checking \verb|t|
  against type \verb;T; except with those occurrences replaced by
  \verb;t₂;.  We also adopt a strong form of Nuprl's \textbf{direct
    computation rules}~\cite{constable+86}: if we have a term $t'$ of
  type $T$ and a proof $t$ that $\{ t' \simeq t''\}$, then we may
  conclude that $t''$ has type $T$ by writing the annotated term
  $\phi\ t\ \mhyph\ t'\{t''\}$, which erases to $t''$.

\item \verb;ι x : T. T';, the dependent intersection type of
  Kopylov~\cite{kopylov03}.  This is the type for terms \verb;t; which
  can be assigned both the type \verb;T; and the type \verb;[t/x]T';,
  the substitution instance of \verb;T'; by \verb;t;.  There are
  constructs \verb;t.1; and \verb;t.2; to select either the \verb;T;
  or \verb;[t.1/x]T'; view of a term \verb;t; of type
  \verb;ι x : T. T';.  We introduce a value of \verb;ι x : T. T'; by
  construct \verb;[t₁, t₂];, where \verb;t₁; has type \verb;T;, \verb;t₂; has type \verb;[t₁/x]T';, and \verb|t₁|
  and \verb|t₂| must have the same erasure (as the intersection type is
  intended as to represent two typings of the same underlying erased term).

\item \verb;∀ x : T. T';, the implicit product type of
  Miquel~\cite{miquel01}.  This can be thought of as the type for
  functions which accept an erased input of type \verb;x : T;, and
  produce a result of type \verb;T';. There are term constructs
  \verb;Λ x. t; for introducing an implicit input \verb;x;, and
  \verb;t -t'; for instantiating such an input with \verb;t';. This
  use of a dash in the notation should not be confused with the uses
  of dash in the notations for $\rho$ and $\phi$ terms, where it is
  just punctuation intended to help separate subexpressions.  The
  implicit arguments exist just for purposes of typing so that they
  play no computational role and equational reasoning happens on terms
  from which the implicit arguments have been erased. Note that
  similar notation is used for quantifications
  $\abs{\forall}{X}{\kappa}{T}$ over types (more generally, type
  constructors), although we use notation $t \cdot T$ instead of $t
  \mhyph T$ to indicate instantiating the quantified type of $t$ with type
  $T$ (that is, for $\forall$-elimination).  These notations
  bind tighter than function space.  If variable $x$ is not free in $T'$,  we write just
  $T \Rightarrow T'$ for $\abs{\forall}{x}{T}{T'}$.
\end{itemize}

\section{HOAS and semantics}
\label{sec:hoas}

The well-known central idea of higher-order abstract syntax (HOAS) is
to encode object-language binders, like $\lambda$ in untyped
$\lambda$-calculus, with meta-language binders.  In a pure type
theory, without introduction of special constructs explicitly for
representation of binders (as in~\cite{miller05}), but rather using
only $\lambda$-abstractions, some puzzles arise:

\begin{enumerate}
\item In pure type theory, all data must be $\lambda$-encoded (e.g., Church-encoded), and
  hence object-language binders would seem automatically to be transformed to $\lambda$-abstractions,
  since all data are.  So it is not clear what could distinguish HOAS from a first-order approach
  to encoding binders.
\item Using $\lambda$-abstractions to encode object-language binders appears too strong, as the
  set of functions even under a strong typing discipline will be much larger than the set of
  weak functions intended to represent the bodies of object-language abstractions.
\end{enumerate}

Washburn and Weirich proposed a solution to (2): use parametric
polymorphism to ensure that, for example, the functions intended to
represent bodies of object-language abstractions cannot pattern-match
on their inputs (which would not correspond to any object-language
abstraction under the usual approach to binding
syntax)~\cite{washburn08}.  They connect their approach to an earlier
work of Sch\"{u}rmann et al., which used modal types to enable similarly restricting
the function space~\cite{schurmann+01}. We will adopt Washburn and Weirich's idea below
(Section~\ref{sec:encoding}), though a twist is required to
obtain a (weakly) initial algebra.

For (1), we may compare with the traditional approach to algebraic
semantics of $\lambda$-calculus (as object language), based on what
are sometimes called environment $\lambda$-models (see Definition 15.3
of~\cite{hindleyseldin}, and cf.~\cite{selinger02}).  Such a model is
a structure $\langle D,\bullet,\interp{-}_{-}\rangle$, where $D$ is a
set of cardinality at least two, consisting of some mathematical objects to be the interpretations of
$\lambda$-terms; $\bullet$ is a binary operation on $D$ intended to
model application; and $\interp{-}_{-}$ is an interpretation function
mapping (object-language) terms $t$ and valuations
$\rho\in\textit{Vars}\to D$ to $D$.  The interpretation function is
required to satisfy various conditions, which suffice to ensure that
the usual equational theory $\lambda\beta$ of $\lambda$-calculus is
sound with respect to $\interp{-}_{-}$: if $\vdash t =_\beta t'$, then
$\interp{t}_\rho = \interp{t'}_\rho$ for any valuation $\rho$.  One of
these conditions, central to soundness of the $\beta$ axiom (scheme), is
that semantic application of the interpretation of a $\lambda$-abstraction
must be the same as evaluating the body with an updated environment:
$\interp{\lam{x}{t}}_\rho \bullet d = \interp{t}_{\rho[x\mapsto d]}$.

If we are looking to universal algebra for ideas on $\lambda$-encoding
HOAS -- as indeed it is profitable to do for encoding first-order
datatypes (see~\cite{wadler90} for a tutorial, or previous work using
Cedille like~\cite{firsov18}) -- we will be misled at this point.  For
environment models presuppose a first-order approach to syntax, so
that they can model instantiation of a $\lambda$-bound variable by
environment update.  And here, even if we functionally encode
valuations, variables, and terms, we will have not achieved anything
beyond usual first-order representations of terms.  To
$\lambda$-encode HOAS, we need a new approach to the semantics of
$\lambda$-calculus that does not use environments.

Categorically, given a endofunctor $F$ on a category $\mathcal{C}$, it
is standard to consider the category of $F$-algebras whose objects are
as $\mathcal{C}$-morphisms from $F\ A$ to $A$ for
$\mathcal{C}$-objects $A$ (the \emph{carrier} of the algebra), and
whose morphisms are $\mathcal{C}$-morphisms $h$ from $A$ to $B$ that
form a commuting square (in $\mathcal{C}$) with the $F\ A$ to $A$ morphisms, and
an $F\ A$ to $F\ B$ morphism derived from $h$.  An
initial algebra is then an initial object in this category, for which
various appealing properties can be proved, in particular that
its carrier $C$ is the least carrier isomorphic to $F\ C$.  From such
developments induction principles are then readily derived.  The
difficulty with HOAS is that the type scheme $F$ one wishes to use is
not a functor, due to a negative occurrence of $X$ in $F\ X$.

\section{An encoding of lambda-terms in Cedille}
\label{sec:encoding}

The basic Church-encoding of inductive types can be carried out in a
type theory like Cedille's, following the categorical perspective.
Given a functorial type scheme $F$, define (within the type theory)
the type $\textit{Alg} \cdot A$ for algebras over type $A$ as $F\ A
\to A$ (recall that in Cedille we use center dot for applying an expression
to a type).  Then the carrier $C$ of a weakly initial algebra has type
$\abs{\forall}{A}{\star}{(F \cdot A \to A) \to A}$.  In the following
discussion, let us write $C_A$ for the type $(F \cdot A \to A) \to A$.
As an example of the definition of $C$: if $F$ is the functor for the
type of natural numbers (and allowing ourselves
infix notation for sum and later product types, and $1$ for unit type), we obtain the type
$\abs{\forall}{A}{\star}{(1 + A \to A) \to A}$ (let
us abbreviate this \textit{Nat}), which is isomorphic to the usual
type $\abs{\forall}{A}{\star}{A \to (A \to A) \to A}$ for
Church-encoded natural numbers.  The main effort is then to define the
algebra itself (not just its carrier), which in general must have type
$\textit{Alg} \cdot C$.  In the case of \textit{Nat}, we need
something of type $\textit{Alg} \cdot \textit{Nat}$, which is easily
obtained: from $1 + \textit{Nat}$ return Church-encoded zero in the
first case, and Church-encoded successor of the given \textit{Nat}
in the second.

\subsection{Starting from Washburn and Weirich}

The approach by Washburn and Weirich, which is not (directly) based on this
perspective, does not allow definition of this algebra.  Their
separate definitions of constructor for object-language
$\lambda$-abstractions and applications can be seen in our terms
as constituting, for the functor $F$ for $\lambda$-terms (which is $\abs{\lambda}{X}{\star}{(X \to X) + (X \times X)}$),  a function of type
$\abs{\forall}{A}{\star}{F \cdot C_A \to C_A}$.
But this is not the type needed for the weakly initial algebra, which
instead should be $\abs{\forall}{A}{\star}{F\cdot C \to C}$.  Without
a definition of a weakly initial algebra, there is no hope, on the
categorical perspective, to define an initial algebra with induction
principle (nor is this claimed in~\cite{washburn08}).

But we may still make use of the basic insight of Washburn and Weirich
that parametricity can be used to restrict the function spaces
intended to represent bodies of object-language abstractions.  To
simplify the discussion (and Cedille code), we consider from here on a
reduced syntax of $\lambda$-terms that omits applications.  So one may
only form terms of the form $\lam{x_1}{\cdots{\lam{x_n}{y}}}$ (and
closed terms require $y\in\{x_1,\ldots,x_n\}$).  This reduced syntax
focuses attention on binding and variable occurrences; adding
applications back in should be completely straightforward.

To return to parametricity: what should be the type of a function
\verb|lam| constructing the encoding of an object-language
$\lambda$-abstraction?  The more fundamental question is, what should
the form \textit{Alg} of algebras be, which will allow construction of
a weakly initial algebra $\textit{Alg} \cdot \textit{Trm}$, where
\textit{Trm} is the desired carrier for encodings of $\lambda$-terms
(without applications)?  It is almost immediately clear that we cannot
use the same notion of algebra as for the Church encoding.  The type
scheme $F$ (it is not a functor) in question is simply $X \to X$, and
thus to inhabit $\textit{Alg} \cdot \textit{Trm}$ we would have to
construct a (meta-language) term of type $(\textit{Trm} \to
\textit{Trm}) \to \textit{Trm}$ (corresponding to $F \cdot C \to C$ in
our general discussion above), and this seems to be impossible.

Drawing inspiration from Selinger's idea of adjoining indeterminates
to an algebra to represent free variables~\cite{selinger02}, let us
think of a binder as introducing a new constructor for the
\textit{Trm} datatype.  So an algebra should be given, for an encoded
lambda abstraction, not just a subterm for the body, but rather a
subterm possibly using a new constructor.  We use parametric
polymorphism to enforce that this binder is abstract.  So we would
like to give our $X$-algebras a function $f$ of type
$\abs{\forall}{Y}{\star}{Y \to \textit{Trm}_Y}$, and obtain from the
algebra then a value of type $X$.  Note that this requires some form
of recursive type so that the type for algebras for \textit{Trm} can
reference \textit{Trm}.  As will be described in a future work (but
see also~\cite{firsov18b}), these are derivable in Cedille.  We elide
calls to fold and unfold these in the following.  The
(candidate) weakly initial algebra would then have type
\begin{equation}
  \label{tp1}
  (\abs{\forall}{Y}{\star}{Y \to \textit{Trm}_Y}) \to \textit{Trm}
  \end{equation}

But there is a problem with this definition.  A requirement we should
impose for the encoding of any datatype is that elements of the
datatype can be built up by successive applications of the
constructors of the datatype (as $3$ can be built by three
applications of the successor constructor to zero).  But if we use
Type~\ref{tp1}, we will not be able to represent object-language
$\lambda$-terms like $\lam{x}{\lam{y}{x}}$.  For Type~\ref{tp1}
requires that the body of the abstraction construct a $\textit{Trm}_Y$
from a $Y$, where $Y$ is abstract.  So the representation of
$\lam{y}{x}$ is not well-typed, because $x$ has some first abstract
type $Y$, while $y$ has a second $Z$, and the body requires a
$\textit{Trm}_Z$.  There is no way to convert $x$ of type $Y$ to $Z$
to embed in a $\textit{Trm}_Z$.

\subsection{A solution using Kripke function spaces}
\label{sec:kripke}

Seen as just considered, we need a way to embed the type of some outer
encoded binder into the types of inner ones.  This is quite
reminiscent of the Kripke semantics for intuitionistic logic, where
implication is interpreted as a modal operator: for $T\to T'$ to be
true at the current world $w$, it must be the case that for all future
worlds $w'$ where $T$ holds, $T'$ also holds.  An $X$-algebra needs
the ability to move the body of the encoded $\lambda$-abstraction to
any world reachable from $X$.  To make the structure of the positive-recursive
type more clear, let us first define a notion like $C_A$ above, but where the
notion of algebra is also a parameter:
\[
  \textit{Trmga} = \abs{\lambda}{\textit{Alg}}{\star \to \star}{\abs{\lambda}{X}{\star}{\textit{Alg} \cdot X \to X}}
\]
We may then give the following positive-recursive definition of algebra:
\begin{equation}
  \textit{Alg} = (\abs{\forall}{Y}{\star}{(X \to Y) \to Y \to \textit{Trmga} \cdot \textit{Alg} \cdot Y}) \to X
  \end{equation}
What we are terming \emph{Kripke function space} rooted at $X$ is a type of the form $\abs{\forall}{Y}{\star}{(X \to Y) \to T}$.
It is the type for functions that can be moved to any type $Y$ reachable from $X$.

This is not the final definition of algebra, though, because as formulated so far, there is no support for iteration.  So the
encoding would be more like a Scott encoding than a Church encoding (see~\cite{stump16} for a comparison).  To support iteration,
the algebra must be given a way to evaluate the value of type $\textit{Trmga} \cdot \textit{Alg} \cdot Y$ returned by its input.
For this, we use Mendler's technique of polymorphically abstracting problematic type occurrences, to allow an algebra to take
in a type-abstracted version of itself~\cite{mendler91}.
\begin{equation}
\begin{array}{lll}
\textit{Alg} & = & \abs{\forall}{\textit{Alga}}{\star\to\star}
       {(\abs{\forall}{Y}{\star}{(X \to Y) \to Y \to \textit{Trmga} \cdot \textit{Alga} \cdot Y})}\\
\ & \ & \textit{Alga}\cdot X \to  \\
\ & \ & (Cast2 \cdot Alg \cdot Alga) \Rightarrow \\
\ & \ & X
\end{array}
  \end{equation}
Here, we have introduced a universal quantification over the type \textit{Alga} of
algebras (one may think of these as \emph{algebra candidates},
similar to Girard's reducibility candidates).  This allows an algebra to be
given an input of type $\textit{Alga} \cdot X$; with just $\textit{Alg}\cdot X$ this
would not be possible as it occurs at a negative position in the recursive definition of \textit{Alg}.
The final input to an algebra is a second-order cast from \textit{Alg} to \textit{Alga}.  Eliding the details,
this allows us to embed any $\textit{Alg}\cdot X$ to an $\textit{Alga}\cdot X$.  This provides the critical
ability for an algebra to interpret encoded terms it is given, possibly using a different
algebra.

Based on this final notion of algebra, we define:
\[
\textit{Trm} \ = \ \abs{\forall}{X}{\star}{\textit{Trmga} \cdot \textit{Alg} \cdot \textit{X}} .
\]
\noindent Evaluation of a term using an algebra is then trivial; terms are functions from algebras to carriers, and so we just apply the term (\textit{t} below) to the algebra (\textit{alg}):
\[
\begin{array}{l}
  \textit{fold}\ :\ \abs{\forall}{X}{\star}{\textit{Alg} \cdot X \to \textit{Trm} \to X} \ = \
  \absu{\Lambda}{X}{\absu{\lambda}{alg}{\absu{\lambda}{t}{t\ \textit{alg}}}}.
\end{array}
\]

More interestingly, we may now define the following algebra
with carrier \textit{Trm}, which we will prove below (Section~\ref{sec:win}) is weakly initial:
\[
\begin{array}{lll}
  \textit{lamAlg} : \textit{Alg}\cdot\textit{Trm}& = & \Lam{\textit{Alga}}{\lam{f}{\Lam{\textit{emb}}{\lam{\textit{talg}}{\ }}}} \\
  \ & \ & \Lam{X}{\lam{\textit{alg}}{\textit{alg} \cdot \textit{Alga}\ (\Lam{Y}{\lam{mx}{f \cdot Y\ (\lam{t}{\textit{mx}\ (t\ \textit{alg})})}})\ \mhyph\textit{emb}\ (\textit{cast2}\ \mhyph\textit{emb}\ \textit{alg})}}.
\end{array}
\]
\noindent All the components discussed above are required here.  We use the
ability to change algebras to invoke \textit{alg} at abstract type
\textit{Alga}, and to make use of \textit{alg} rather than
\textit{talg}.  We can notice that \textit{talg} is not even used
(note that in the application $\textit{mx}\ (t\ \textit{alg})$, we
have $t$ applied to \textit{alg}, not \textit{talg}).  So rather than
recursing through the body of the encoded $\lambda$-abstraction as
given by $f$ using the algebra which is being given to
\textit{lamAlg}, \textit{lamAlg} instead switches algebras to use the
one being given to the \textit{Trm} which it (\textit{lamAlg}) is
being asked to produce.  A cast changes the type of
\textit{alg} to the instance $\textit{Alga}\cdot X$ of the abstracted
algebra.

For use in nested construction of terms, the following variant of \textit{lamAlg}
is needed:
\[
\begin{array}{l}
\textit{lam} : \abs{\forall}{X}{\star}{(\abs{\forall}{Y}{\star}{(X \to Y) \to Y \to \textit{Trmga}\cdot \textit{Alg}\cdot Y}) \to \textit{Trmga}\cdot \textit{Alg}\cdot X} \\
\hspace{1cm} = \Lam{X}{\lam{f}{\lam{alg}{\textit{alg} \cdot \textit{Alg}\ f\ \mhyph(\textit{castId2} \cdot \textit{Alg})\ \textit{alg}}}}
\end{array}
  \]
  The difference from \textit{lamAlg} is that here the Kripke function space is rooted at any type $X$, where \textit{lamAlg} is rooted at \textit{Trm}.  Quantifying over the root of the Kripke function space allows nested applications of \textit{lam}, as in the encoding of the second-projection function (first defining a convenience function \textit{place}):
  \[
  \begin{array}{lll}
    \textit{place} : \abs{\forall}{X}{\star}{X \to \textit{Trmga} \cdot \textit{Alg} \cdot X} & = & \Lam{X}{\lam{x}{\lam{\textit{alg}{x}}}} \\
    \\
    \textit{proj2} : \textit{Trm} & = &
      \Lam{O}{\textit{lam}\ (\Lam{X}{\lam{\textit{mo}}{\lam{x}{\ }}}} \\
      \ & \ & \hspace{.7cm}       \textit{lam}\ (\Lam{Y}{\lam{\textit{mx}}{\lam{y}{\textit{place}\ (\textit{mx}\ x)}}}))
  \end{array}
  \]
  Notice how the outer meta-language bound variable $x$ is used inside the (meta-language) binding of $y$, using \textit{mx} to move it from $X$ to $Y$.

  The inspiration of Kripke semantics for semantics of lambda calculus
  may also be found in works like Mitchell and
  Moggi's~\cite{mitchell+1991}.  There, explicit environments
  are used to interpret terms, and so the semantics fails to be a
  suitable basis for a higher-order encoding, for the reasons
  discussed above.

\section{Haskell listing}
\label{sec:hs}

The above development actually does not make use of the special
features of Cedille beyond (derivable) positive-recursive types.  In
fact, it can be carried out in any language supporting impredicative
quantification and positive recursive types, such as Haskell
(impredicativity has to be mediated by inductive datatypes in a
certain way, but is essentially present).  To aid the reader more
familiar with Haskell than Cedille, Figure~\ref{fig:hs} gives a
Haskell listing of the functions discussed above.  This requires
Haskell \verb|LANGUAGE| extensions \verb|KindSignatures|,
\verb|ExplicitForAll|, and \verb|RankNTypes|.  Some uses of implicit
function space in the Cedille code have been converted to the regular
(explicit) function spaces of Haskell.  Impressively, Haskell's type
inference is powerful enough to allow us to avoid type annotations
except for marking the places where universal generalization (with
constructors \texttt{MkAlg} and \texttt{Trm}) and instantiation (with
eliminators \texttt{unfoldAlg} and \texttt{unfoldTrm}) occur.

\begin{figure}
\small
\begin{verbatim}
module WeaklyInitialHoas where

type Trmga alg x = alg x -> x

newtype Alg x =
  MkAlg { unfoldAlg :: forall (alga :: * -> *) .
                        (forall (y :: *) . (x -> y) -> y -> Trmga alga y) ->
                        (forall (z :: *) . Alg z -> alga z) ->
                        alga x -> x}
newtype Trm = MkTrm { unfoldTrm :: forall (x :: *) . Alg x -> x}

fold :: Alg a -> Trm -> a
fold alg t = unfoldTrm t alg

lamAlg :: Alg Trm
lamAlg = MkAlg (\ f embed talg -> 
           MkTrm (\ alg ->
             unfoldAlg alg (\ mx -> f (\ t -> mx (unfoldTrm t alg)))
               embed (embed alg)))

lam :: forall (x :: *) . 
       (forall (y :: *) . (x -> y) -> y -> Trmga Alg y) -> Trmga Alg x
lam = \ f alg -> unfoldAlg alg f (\ x -> x) alg

place :: forall (x :: *) . x -> Trmga Alg x
place = \ x -> \ alg -> x
\end{verbatim}
\caption{Haskell definitions for the Cedille code above}
\label{fig:hs}
\end{figure}

\section{Examples}

Based on the Haskell implementation of Figure~\ref{fig:hs}, let us
consider several examples of algebras.  For testing, we will use the
following simple term, representing
$\absu{\lambda}{x}{\absu{\lambda}{y}{x}}$:
\begin{verbatim}
test :: Trm
test = MkTrm (lam (\ mo x ->
               lam (\ mx y -> place (mx x))))
\end{verbatim}

\subsection{Size}

\begin{figure}
\begin{verbatim}
sizeAlg :: Num a => Alg a
sizeAlg = MkAlg (\ f embed alg -> 1 + f id 1 alg)
\end{verbatim}
\caption{An algebra for the size of a term}
\label{fig:size}
\end{figure}

Figure~\ref{fig:size} gives an algebra for computing the size of a
term.  The algebra is given the body \texttt{f}, the embedding
\texttt{embed} from \texttt{Alg} to abstract type \texttt{Alga} (which
is not needed for this example), and the algebra itself under the
abstract type.  Interpreting a $\lambda$-abstraction (as is done by
this and all algebras) is done by interpreting the body using
\texttt{alg}, where \texttt{1} is given as the value to use for the
bound variable.  The use of the \texttt{id} (identity function in
Haskell) is to map trivially from the carrier of the algebra to the
type at which we are interpreting the body, namely also the carrier.

\noindent Interpreting our test term with \texttt{sizeAlg} gives us the following
interaction using \verb|ghci|:
\begin{verbatim}
*WeaklyInitialHoas> fold sizeAlg test
3
\end{verbatim}
This is as expected, size we count one for each $\lambda$ and then one
for the use of the variable $x$.

\subsection{Converting to strings}

\begin{figure}
\begin{verbatim}
vars :: Int -> [String]
vars n = ("x" ++ show n) : vars (n + 1)

printTrmAlg :: Alg ([String] -> String)
printTrmAlg = 
  MkAlg (\ f embed alg vars ->
           let x = head vars in
               "\\ " ++ x ++ ". " ++ f id (\ vars -> x) alg (tail vars))

printTrm :: Trm -> String
printTrm t = fold printTrmAlg t (vars 1)
\end{verbatim}
\caption{Algebra and related functions for converting a term to a string}
\label{fig:print}
\end{figure}

Figure~\ref{fig:print} defines an algebra \verb|printTrmAlg| for use
in converting a term to a string.  The carrier of the algebra is
\verb|[String] -> String|; a term is interpreted as a function from a
stream of variable names (the \verb|[String]|) to a \verb|String|
representation of the term.  The algebra simply peels off the first
name (\verb|x| in the code) from the stream and uses it for the
binding occurrence of the variable.  For any bound occurrences in the
body \verb|f|, the algebra passes to \verb|f| the function
\verb|\ vars -> x| as the interpretation for the variable.  This
function (of type \verb|[String] -> String|) simply discards the
stream of names it is given and returns \verb|x| as the interpretation
(i.e., the string representation) of the bound variable.

For \verb|printTerm|, we fold the algebra over the input term, and then
apply the resulting function to the simple stream of variable names \verb|vars 1|.
For our test term, we can observe the following result with \verb|ghci|:

\begin{verbatim}
*WeaklyInitialHoas> putStrLn (printTrm test)
\ x1. \ x2. x1
\end{verbatim}

\subsection{Converting to de Bruijn notation}

\begin{figure}
\begin{verbatim}
data Dbtrm = Lam Dbtrm | Var Int deriving Show

toDebruijnAlg :: Alg (Int -> Dbtrm)
toDebruijnAlg = MkAlg (\ f embed alg -> \ v ->
                   let v' = v + 1 in Lam (f id (\ n -> Var (n - v')) alg v'))
\end{verbatim}
\caption{An algebra for converting to de Bruijn notation}
\label{fig:db}
\end{figure}

Figure~\ref{fig:db} defines a datatype \verb|Dbtrm| for untyped
$\lambda$-terms in de Bruijn notation (without application, similarly
to our running example). The figure also defines an algebra for
converting a term to a \verb|Dbtrm|.  More precisely, the carrier of
the algebra is \verb|Int -> Dbtrm|; the algebra converts a term to a
function which takes in the number \verb|v| to use as the current
depth of nesting within $\lambda$-abstractions.  The algebra
interprets the body with the successor nesting depth \verb|v'|.  It
supplies the function \verb|\ n -> Var (n - v')| for the
interpretation of the bound variable.  This function takes in the
current depth \verb|n| and subtracts off \verb|v'|, which is one plus
the depth at which the binding occurrence of the variable was
encountered (subtracting one ensures that the starting de Bruijn index
is zero).  For the test term, we confirm the expected result with
\verb|ghci|:

\begin{verbatim}
*WeaklyInitialHoas> fold toDebruijnAlg test 1
Lam (Lam (Var 1))
\end{verbatim}

\section{Weak initiality of lamAlg}
\label{sec:win}

\begin{figure}
\begin{verbatim}
IsHom : Π X1 : ★ . (Alg · X1) ➔
        Π X2 : ★ . (Alg · X2) ➔
        Π h : X1 ➔ X2 . ★ =
  λ X1 : ★ . λ alg1 : Alg · X1 .
  λ X2 : ★ . λ alg2 : Alg · X2 .
  λ h : X1 ➔ X2 .
    ∀ Alga : ★ ➔ ★ .
    ∀ f : ∀ Y : ★ . (X1 ➔ Y) ➔ Y ➔ Trmga · Alga · Y .
    ∀ c : Cast2 · Alg · Alga .
    { h (alg1 f alg1) ≃ alg2 (λ mx . f (λ a . mx (h a))) alg2 }.
\end{verbatim}
\caption{Definition of homomorphism}
\label{fig:hom}
\end{figure}

We would now like to consider the above development from a categorical
perspective, as is standard for simpler classes of inductive datatypes
like those arising from polynomial functors (see~\cite{bird+97} for a
summary in service of functional programming).  Given two algebras
\texttt{alg1} and \texttt{alg2} with carriers \verb|X1| and \verb|X2|,
we must first define what it means for a function \verb|h : X1 ➔ X2| to
be a homomorphism from the first algebra to the second.  The definition
is given, in Cedille notation, in Figure~\ref{fig:hom}.  It states that
such an \verb|h| is a homomorphism iff for all components required by \texttt{alg1} -- that is, for all algebra candidates \verb|Alga|, bodies \verb|f|, and embeddings \verb|c| -- the following equation holds:

\begin{verbatim}
{ h (alg1 f alg1) ≃ alg2 (λ mx . f (λ a . mx (h a))) alg2 }
\end{verbatim}
\noindent
This is an adaptation of the usual commutation condition one desires
for homomorphisms.  It says that applying the homomorphism and then
\verb|alg1| (to \verb|f|) is the same as applying \verb|alg2| to a
modified version of \verb|f|, which applies \verb|h| internally.

Using this definition of homomorphism, we can prove (in Cedille) the
theorems shown in Figure~\ref{fig:algthms}. The first says that
\verb|λ x . x| is a homomorphism from any algebra to itself.  The
second states that homomorphisms compose.  The third is the main
result of the paper. It states that \verb|lamAlg| (defined in
Section~\ref{sec:kripke} above) is a weakly initial algebra: for any
algebra \verb|alg|, the function \verb|fold alg| is a homomorphism
from \verb|lamAlg| to \verb|alg|.  These theorems have short simple
proofs, as one would anticipate.

\begin{figure}
\begin{verbatim}
IdHom : ∀ X : ★ . ∀ alg : Alg · X .
        IsHom · X alg · X alg (λ x . x)

ComposeHom : ∀ X1 : ★ . ∀ alg1 : Alg · X1 .
             ∀ X2 : ★ . ∀ alg2 : Alg · X2 .
             ∀ X3 : ★ . ∀ alg3 : Alg · X3 .
             ∀ h1 : X1 ➔ X2 .
             ∀ h2 : X2 ➔ X3 .
             IsHom · X1 alg1 · X2 alg2 h1 ➔
             IsHom · X2 alg2 · X3 alg3 h2 ➔
             IsHom · X1 alg1 · X3 alg3 (λ x . h2 (h1 x))

foldHom : ∀ X : ★ . ∀ alg : Alg · X .
IsHom · Trm lamAlg · X alg (fold alg)
\end{verbatim}
\caption{Algebras form a category with \texttt{lamAlg} as a weakly initial object}
\label{fig:algthms}
\end{figure}

\section{Conclusion}
\label{sec:param}

In this paper, we have seen how to derive a weakly initial algebra for
a very simple datatype using higher-order abstract syntax.  The
crucial next step of this work in progress is to extend the
development to derive an initial (not just weakly initial) algebra,
for the \textit{Trm} datatype.  The strategy I am following for this
is to form a dependent intersection of \textit{Trm} as defined above
with a statement of unary parametricity~\cite{reynolds83}.  It should
be possible to do this for any type (and hence for \textit{Trm}), as
studied by Bernardy and Lasson~\cite{bernardy11}).  And with a
reflection principle that can hopefully be baked into the definition
of the datatype, unary parametricity implies induction.  The next
bigger step is to try to give a generic development of induction with
HOAS, for any type scheme satisfying certain (as yet to be delineated)
restrictions.  The final goal is to extend Cedille's datatype
notations to allow HOAS, and elaborate those notations down to the
generic version of induction for HOAS.

\textbf{Acknowledgments.} I gratefully acknowledge NSF support under
award 1524519, and DoD support under award FA9550-16-1-0082 (MURI
program). Many thanks to the anonymous LFMTP '19 reviewers for very
helpful feedback, incorporated into the final version, including the
suggestion to include conversion to de Bruijn notation as an example.
AMDG.

\bibliographystyle{eptcs}
\bibliography{paper}

\end{document}